\documentclass[12pt,preprint]{aastex}
\usepackage{multirow}
 \newcommand{\beq}{\begin{equation}}
 \newcommand{\eeq}{\end{equation}}
 \newcommand{\beqn}{\begin{eqnarray}}
 \newcommand{\eeqn}{\end{eqnarray}}
 \newcommand{\non}{\nonumber}
 \newcommand{\no}{\noindent}

\newcommand{\pa}{\partial}
\newcommand{\cl}{\centerline}

\begin{document}
\title{Galaxies with Supermassive Binary Black Holes:\\
(I) A Possible Model for the Centers of Core Galaxies} 

\author{Ing-Guey Jiang$^{1}$ and Li-Chin Yeh$^{2}$}

\affil{
{$^{1}$ Department of Physics and Institute of Astronomy,}\\
{ National Tsing-Hua University, Hsin-Chu, Taiwan}\\ 
{$^{2}$ Department of Applied Mathematics,}\\
{ National Hsinchu University of Education, Hsin-Chu, Taiwan} 
}

\email{jiang@phys.nthu.edu.tw}

\begin{abstract} 

The dynamics of galactic systems with central binary black holes
is studied. The model is a modification from the
restricted three body problem, in which a galactic potential is 
added as an external potential.  Considering
the case with an equal mass binary black holes,  
the conditions of existence of equilibrium points,
including Lagrange Points and 
additional new equilibrium points, i.e. Jiang-Yeh Points,
are investigated. A critical mass is discovered to be 
fundamentally important. That is, Jiang-Yeh Points exist  
if and only if the galactic mass is larger than the
critical mass. The stability analysis is performed for 
all equilibrium points. The results that  
Jiang-Yeh Points are unstable could lead to the
core formation in the centers of galaxies. 
\end{abstract}

\newpage
\section{Introduction}

A modification of the restricted three body 
problem was first studied in Chermnykh (1987). 
The angular velocity variation of the system was considered
and  the stability of solutions near equilibrium points
was investigated.
Extending from the work of Chermnykh (1987),
Papadakis (2004) studied the symmetric motions
near the three collinear equilibrium points,
Papadakis (2005a) investigated 
the stability of the periodic orbits and explored
the network of the orbital families, 
Papadakis (2005b) provided
the analytic determination of the initial conditions of the long- 
and short-period Trojan families around the equilibrium points
of the Sun-Jupiter system. 

Then, Jiang \& Yeh (2006) and Yeh \& Jiang (2006) 
investigated the existence of new equilibrium points
of a system with binary masses and a disk as a modification of
the restricted three body problem, which was hereafter called  
Chermnykh-like problem for convenience 
(Jiang \& Yeh 2006, Yeh \& Jiang 2006).  
Moreover, Kushvah (2008a, 2008b, 2009) studied the linear stability 
of equilibrium points of the generalized
photo-gravitational problem. The trajectories and Lyapunov 
characteristic exponents are also calculated and studied in 
Kushvah (2011a, 2011b).

Furthermore, the recent rapid development of 
the discoveries of extra-solar planetary 
systems has triggered many interesting theoretical work
(see Jiang \& Ip 2001, Ji et al. 2002, 
Jiang et al. 2003, Jiang \& Yeh 2004a, 
Jiang \& Yeh 2004b, Jiang \& Yeh 2004c,
Jiang \& Yeh 2007,
Jiang \& Yeh 2009, 
Jiang \& Yeh 2011), statistics work 
(see  Jiang et al. 2006, 2007, 2009, 2010), 
and also the following-up observational work (Jiang et al. 2013)
in this field.
In fact, some results of the Chermnykh-like problem 
could have important applications on 
the dynamical problems of planetary systems
(Papadakis 2005b, Kushvah 2011a).

On the other hand, the dynamical evolution of galaxies
is another field that Chermnykh-like problems could 
be used as theoretical models. 
It is now generally believed that 
nearly every galaxy hosts a supermassive black hole at the center.
It is also believed that most massive
elliptical galaxies were formed by merging of existing galaxies.
Thus, it is possible that many luminous elliptical galaxies 
might host binary black holes.
The orbital motion of stars near the binary black hole at 
the centers of galaxies is a dynamical system that 
Chermnykh-like problems could act as a good model.
The possible orbits of these stars would
be related with the density profile near the cores
of galaxies.
The brightness near the core of galaxies does 
show some signature which might be related with 
the dynamical evolution of stars near the binary 
black hole.  

For example, Lauer et al. (2007) 
investigated the cores of early-type galaxies.
The detail brightness profiles are systematically classified. 
However, it is unclear how the cores were formed and how
core properties are related with the dynamical evolution of 
these galaxies.
Kandrup et al. (2003) studied the stellar distribution
under the influence of a binary black hole and 
a galactic potential. A systematic transport of 
stars near the black holes out to larger radii
was investigated and the resulting 
stellar density profile was used to explain the 
observed profile of NGC 3706.
   
In order to seek a possible picture to explain the 
core formation of these early-type galaxies, 
we here employ a model with a supermassive binary black hole 
in a galaxy. 
To analytically understand the dynamical
properties of this system, we first investigate the 
existence of equilibrium points.
In addition to the existence of Lagrange points, 
the conditions under which other equilibrium points
would exist are also investigated. 
After that, the stability of all equilibrium points
is discussed. 

The model is given in Section 2,
the analytic results of equilibrium points are in Section 3,
the bifurcation diagrams and locations of equilibrium points 
are shown in Section 4, the stability analysis is in Section 5,
and finally the summary is in
Section 6.
 
\section{The Model}

We study the dynamical motion of a star near the center of a galaxy.
The star, considered as a test particle,
is influenced by 
the galactic potential and the gravitational force from 
the central supermassive binary black hole
(SBBH). Depending on the mass ratio between the galaxy and 
SBBH, the separation of SBBH, and also the central concentration
of the galactic density profile, the central region with SBBH
could be dominated by the SBBH, and thus the galactic potential
could be ignored approximately. 
In this case, when the SBBH is moving on
circular orbits,  
the star's equations of motion are exactly
as in a restricted three body problem:
 
(Murray \& Dermott 1999, Jiang \& Yeh 2006)
\beq\left\{
\begin{array}{ll}
&\frac{d\bar{x}}{d\bar{t}}=\bar{u} \\
&\frac{d\bar{y}}{d\bar{t}}=\bar{v}  \\
&\frac{d\bar{u}}{d\bar{t}}=2\bar{n}\bar{v}-\frac{\pa U^{\ast}}{\pa \bar{x}} \\
& \frac{d\bar{v}}{d\bar{t}}=-2\bar{n}\bar{u}-\frac{\pa U^{\ast}}{\pa \bar{y}},\\\end{array} \right. \label{eq:3body1}
\eeq
where $\bar{n}$ is the central binary's angular velocity,
\beq
U^{\ast}=-\frac{\bar{n}^2}{2}(\bar{x}^2+\bar{y}^2)-\frac{G\bar{m_1}}{\bar{r}_1}
-\frac{G\bar{m_2}}{\bar{r}_2}.
\label{eq:u_ast}
\eeq
The Jacobi integral of the above system is 
\beq
{\bar C_J} = -{\bar u}^2  -{\bar v}^2 - 2  U^{\ast}.
\label{eq:CJ_old} 
\eeq

The solution and stability of the above system 
are well studied (Murray \& Dermott 1999). 
In this paper, we are interested in the case with 
equal mass SBBH, thus we set the mass 
$\bar{m}= \bar{m_1} = \bar{m_2} $ and 
the binary has a separation $2\bar {R}$. 
Moreover,

$$ \bar{r}=\sqrt{\bar{x}^2+\bar{y}^2},   
\bar{r}_1=\sqrt{(\bar{x}+\bar{R})^2+\bar{y}^2}, \quad {\rm and}\,
\quad \bar{r}_2=\sqrt{(\bar{x}-\bar{R})^2+\bar{y}^2}. $$
The equations of motion are now written as:
\beq \left\{
\begin{array}{ll}
&  \frac{d\bar{x}}{d\bar{t}}=\bar{u}  \\
& \frac{d\bar{y}}{d\bar{t}}=\bar{v}  \\
& \frac{d\bar{u}}{d\bar{t}}=2\bar{n}\bar{v} +\bar{n}^2\bar{x}-\frac{G\bar{m}(
\bar{x}+\bar{R})}{\bar{r}_1^3}-\frac{G\bar{m}(\bar{x}-\bar{R})}
{\bar{r}_2^3}\\
& \frac{d\bar{v}}{d\bar{t}}=-2\bar{n}\bar{u}+\bar{n}^2 \bar{y}-\frac{G\bar{m}\bar{y}}{\bar{r}_1^3}-\frac{G\bar{m}
\bar{y}}{\bar{r}_2^3}. \\
\end{array}  \right. \label{eq:3body2}
\eeq

As the procedure in Yeh, Chen, \& Jiang (2012), 
it is better to do
non-dimensionalization for the above system. This
would not only simplify equations, but also 
make it more convenient when we apply  
mathematical models 
on real physical systems. 
When  non-dimensional variables are set as
$$x=\frac{\bar{x}}{L_0}, y=\frac{\bar{y}}{L_0},R=\frac{\bar{R}}{L_0}, \,\,
 t=\frac{\bar{t}}{t_0},  m=\frac{\bar{m}}{m_0}, u=\frac{\bar{u}}{u_0},
v=\frac{\bar{v}}{u_0}, n_b={t_0}{\bar{n}}
$$
and some of the above parameters are further assumed as  
$u_0=\frac{L_0}{t_0}$ and $Gm_0=\frac{L_0^3}{t_0^2}$,
then System (\ref{eq:3body2}) can be rewritten as
\beq \left\{
\begin{array}{ll}
&  \frac{dx}{dt}=u  \\
& \frac{dy}{dt}=v  \\
& \frac{du}{dt}=2n_bv +n^2_b x-\frac{m(x+R)}{r_1^3}-\frac{m(x-R)}
{r_2^3} \\
& \frac{d v}{dt}=-2n_bu+n^2_b y-\frac{my}{r_1^3}-
\frac{my}{r_2^3},
\end{array}  \right. \label{eq:3body_nondim}
\eeq
where $n_b^2=\frac{m}{4R^3}$, $r^2=x^2+y^2$, $r_1^2=(x+R)^2+y^2$, and
$r_2^2=(x-R)^2+y^2$.

The above are the equations when the galactic potential is 
ignored in our system.
However, in order to generally consider all different 
masses of the galaxy and SBBH for the study of  
stellar dynamics 
near SBBH in the region of a galactic center,
the galactic potential shall be included.
The Nuker law, which was originally used to 
characterize the inner light distributions of
galaxies (Lauer et al. 1995), is employed to set the
overall galactic dark-matter plus luminous density 
distribution here.

The Nuker law is expressed as
\beq
\rho(\bar{r})=\rho_c\left(\frac{\bar{r}}{r_b}\right)^{-\gamma}
\left\{1+\left(\frac{\bar{r}}
{r_b}\right)^{\alpha}\right\}^{\frac{\gamma-\beta}{\alpha}}\,,
\label{rho_bar_r}
\eeq
where $\rho_c, \alpha, \beta, \gamma$ are constants.
The scale length $r_b$ is called the break radius.
When $\bar{r} \gg r_b$, the above becomes a power law with index
$\beta$. For $\bar{r} \ll r_b$, 
$\gamma$ is the inner cusp slope 
of this density profile
as $\bar{r}$ approaches to zero. 
Thus, the break radius, $r_b$,  marks a transition from the 
outer power law with index $\beta$
to inner cusp with index $\gamma$.  
Usually, the slope of outer profile is steeper than the central 
cusp, so that $\beta$ is larger than $\gamma$.
Core galaxies are those that exhibit clear break radius
but power-law galaxies do not.
Both types are studied and classified in Lauer et al. (2007).
The properties of central profiles of these two types of galaxies
might be related with the general structures and formation 
histories of galaxies and thus could be fundamentally important.
Because the break radius is so important, 
we set our length scale $L_0=r_b$ here. 
In other words, we regard $r_b$ as a typical length scale
in our system and we have  
\beq 
r=\frac{\bar{r}}{r_b}
\eeq 
and a dimensionless Nuker law
\beq
\rho(r)=\rho_c\left( r\right)^{-\gamma}
\left\{1+\left( r \right)^{\alpha}\right\}^{\frac{\gamma-\beta}{\alpha}}\,.
\label{rho}
\eeq

According to the observational results and classifications 
in Lauer et al. (2007), the brightness of galaxies 
present different values of $\alpha, \beta$ and $\gamma$
in the above formula.  As a first step, in this paper, we choose
 $\alpha=2$, $\beta=4$ and $\gamma=0$ (Kandrup et al. 2003).
The mass of the galaxy upto a radius $r$ is
\beq
M(r)= 4\pi \int^r_0 r^2\rho(r)dr
=2\pi\rho_c\left\{\tan^{-1}r-\frac{r}{1+r^2}\right\}.
\eeq
Note that the mass, potential, radius, and 
normalization constants etc. are all expressed
as dimensionless quantities hereafter for convenience.  
Obviously, $M(r)$ does not include the mass of SBBH.
Thus, if the galactic total mass is $M_g$, 
then we have $\rho_c=\frac{M_g}{\pi^2}$.

On the other hand, 
the corresponding potential is 
\beq
V(r)=-4\pi\left[\frac{1}{r}\int_0^r \rho(\bar r)\bar r^2d\bar r +\int_r^\infty \rho
(\bar r)\bar rd\bar r\right]=-c\frac{\tan^{-1}r}{r},\label{eq:va1}
\eeq
where $c=\frac{2}{\pi}M_{\rm g}$. 
Moreover, from the potential, we can obtain the 
gravitational force as 
\beq
f_g(r)\equiv -\frac{\pa V}{\pa r}=
c \left\{\frac{1}{(1+r^2)r}-\frac{\tan^{-1} r}{r^2}
\right\}. 
\label{eq:fg}
\eeq
Including this axis-symmetric force into the
system, our equations of motion now become:
\beq \left\{
\begin{array}{ll}
&  \frac{dx}{dt}=u  \\
& \frac{dy}{dt}=v  \\
& \frac{du}{dt}=2nv +n^2 x-\frac{m(x+R)}{r_1^3}-\frac{m(x-R)}
{r_2^3}+\frac{c x}{r} \left\{\frac{1}{(1+r^2)r}-\frac{\tan^{-1} r}{r^2}
\right\}, \\
& \frac{d v}{dt}=-2nu+n^2 y-\frac{my}{r_1^3}-
\frac{my}{r_2^3}+\frac{c y}{r}
 \left\{\frac{1}{(1+r^2)r}-\frac{\tan^{-1} r}{r^2}\right\}.
\end{array}  \right. \label{eq:3body3_non}
\eeq
Note that the length scale $L_0$ of the above equations 
has been set to 
equal to the break radius $r_b$ of 
the galaxy we considered in this system.
Moreover,  
under the influence of the galactic potential, 
the mean motion, i.e.
angular velocity, of each component of SBBH in the above equations 
now becomes  
\beq
n= \left\{ \frac{m}{4R^3}+\frac{1}{R}|f_g(R)|\right\}^{1/2}. 
\label{eq:n2}
\eeq
The corresponding Jacobi integral of this system is, similar as
the one in Jiang \& Yeh (2006),
\beq
C_J =  -u^2 -v^2 + n^2 (x^2+y^2) 
       + \frac {2m}{r_1} + \frac {2m}{r_2}
       + 2c\frac{ {\rm tan}^{-1}(\sqrt{x^2+y^2}) }{\sqrt{x^2+y^2}}. 
\label{eq:Jacobi}
\eeq

\section{The Equilibrium Points}

It is well-known that there are five equilibrium points, i.e. 
Lagrange Points, for the
classical restricted three-body problem.  
In the case that when an additional potential is added,
the results in Jiang \& Yeh (2006) and Yeh \& Jiang (2006)
show that, two more equilibrium points, called 
Jiang-Yeh Points hereafter for convenience, might exist 
under some conditions.

We study the system of equal mass SBBH with galactic potential here.
In this paper, the equilibrium point at the origin (0,0) will be 
named as Lagrange Point 1 (L1), the one at $x$-axis with $x>R$
is named as Lagrange Point 2 (L2),  the one at $x$-axis with $x<-R$
is named as Lagrange Point 3 (L3), the one at $y$-axis with $y>0$
is named as Lagrange Point 4 (L4), the one at $y$-axis with $y<0$
is named as Lagrange Point 5 (L5), the one at $x$-axis with $0<x<R$
is called as Jiang-Yeh Point 1 (JY1), the one at $x$-axis with $-R<x<0$
is called as Jiang-Yeh Point 2 (JY2).
In this section, the existence of the above equilibrium points  
is investigated and the conditions of their existence
will be determined and proved. 

In general, for System
(\ref{eq:3body3_non}), equilibrium points $(x_e,y_e)$
satisfies $A(x_e,y_e)=0$ and $B(x_e,y_e)=0$, where
\beqn
&&A(x,y)= n^2 x-\frac{m(x+R)}{r_1^3}-\frac{m(x-R)}{r_2^3}+\frac{c x}{r}
 \left\{\frac{1}{(1+r^2)r}-\frac{\tan^{-1} (r)}{r^2}\right\},
\label{eq:gf2}\\
& & B(x,y)=n^2 y-\frac{m y}{r_1^3}-\frac{m y}{r_2^3}
+\frac{c y}{r}
 \left\{\frac{1}{(1+r^2)r}-\frac{\tan^{-1}(r)}{r^2}\right\}. \label{eq:gg2}
\eeqn

For convenience, for $y\neq 0$, 
we define
\beqn
h(y)&\equiv& \frac{B(0,y)}{y}=n^2-\frac{2m}{\left[R^2+y^2\right]^{3/2}}
+\frac{c}{|y|}
\left\{\frac{1}{(1+y^2)|y|}-\frac{\tan^{-1}(|y|)}{y^2}\right\}  \non \\
&=&n^2-\frac{2m}{\left[R^2+y^2\right]^{3/2}}
+c\left\{\frac{1}{(1+y^2)y^2}-\frac{\tan^{-1}(y)}{y^3}\right\}
\label{eq:hye}
\eeqn
and then
\beqn
k(x)&\equiv& A(x,0)=n^2x-\frac{m(x+R)}{|x+R|^3}-\frac{m(x-R)}
{|x-R|^3} +\frac{cx}{|x|}\left\{\frac{1}
{(1+x^2)|x|}-\frac{\tan^{-1}(|x|)}{x^2}\right\} \non \\
&=& n^2x-\frac{m(x+R)}{|x+R|^3}-\frac{m(x-R)}{|x-R|^3}+c
\left\{\frac{1}
{(1+x^2)x}-\frac{\tan^{-1}(x)}{x^2}\right\}
 \label{eq:kxe}
\eeqn

Because the proofs of Property 3.1 and 3.2 of Jiang \& Yeh (2006)
does not involve the detail form of the force derived from 
the additional potential, 
the result that the equilibrium points are either on the $x$-axis 
or $y$-axis is valid for our system of equal mass SBBH.
Besides, 
we have the following Remarks for our system here.\\
{\bf  Remark (A):} \\
\no For $y_e \neq 0$,  $y_e$ satisfies  $h(y)=0$, if and only if
$(0,y_e)$  is
the equilibrium point of System (\ref{eq:3body3_non}).\\
{\bf  Remark (B):} \\
\no $x_e$ satisfies $k(x)=0$, if and only if
$(x_e,0)$  is  the equilibrium point of System (\ref{eq:3body3_non}).

Employing the above Remarks, the  
equilibrium points L4 and L5 will be studied in 
Theorem 1, and all the other
equilibrium points on $x$-axis will be investigated in 
Theorem 2, 3, 4.\\ 

\no {\bf Theorem 1: The Existence and Uniqueness of Lagrange Points L4 and L5}\\There is one and only one 
$\bar{y_1}>0$ such that $h(\bar{y_1})=0$, and only one 
$\bar{y_2}<0$ such that $h(\bar{y_2})=0$. 
That is, excluding the origin (0,0) of $x-y$ plane, 
there are two equilibrium points on the $y$-axis, i.e. L4 and L5,
for System (\ref{eq:3body3_non}). 

{\bf (Proof): }  
We define
\beq
P(y)\equiv\frac{2m}{\left[R^2+y^2\right]^{3/2}}-n^2 \label{eq:p1}
\eeq
 \beq
{\rm and} \qquad Q(y)\equiv c\left\{\frac{1}{(1+y^2)y^2}
-\frac{\tan^{-1}(y)}{y^3}\right\},
\label{eq:q1}
\eeq 
so from Eq.(\ref{eq:hye}), we have $h(y)= -P(y)+Q(y). $

We consider the case when $y>0$ first.
From Eqs. (\ref{eq:p1})-(\ref{eq:q1}), we have
\beq
P'(y)= -6my\left[R^2+y^2\right]^{-5/2}<0, \label{eq:dp1}
\eeq
 and
\beq
Q'(y)= c\frac{-3y-5y^3+3(1+y^2)^2\tan^{-1}(y)}
{y^4(1+y^2)^2}
= c\left\{\frac{3(1+y^2)^2g_1(y)}{y^4(1+y^2)^2}\right\}, 
\label{eq:dq1}
\eeq
where $g_1(y)\equiv \tan^{-1}y-\frac{5y^3+3y}{3(1+y^2)^2}$.  Since 
$g_1'(y)=\frac{8y^4}{3(1+y^2)^3}>0$ for $y>0$ and $g_1(0)=0$, we 
have $g_1(y)>0$ for $y>0$. Thus, from Eq.(\ref{eq:dq1}), we know
$Q'(y)>0$ for all $y>0$. Therefore, 
both $P(y)$ and $Q(y)$ are monotonic functions, 
and $h(y)=-P(y)+Q(y)$ is a monotonically increasing function.

On the other hand, we have 
\beq
\lim_{y\to 0}P(y)= \frac{2m}{R^3}-n^2,
\label{eq:P_lim_0}
\eeq
and 
\beq
\lim_{y\to \infty}P(y)=-n^2.
\label{eq:P_lim_inf}  
\eeq
Moreover, using L'Hopital Rule
on Eq.(\ref{eq:q1}), we have
\beqn
\lim_{y\to 0}Q(y)&=&\lim_{y\to 0}\frac{c}{y^3}\left(\frac{y}{1+y^2}-
\tan^{-1} y\right) \non \\
&=& \lim_{y\to 0}\frac{-2cy^2}{3y^2(1+y^2)^2}=
-\frac{2c}{3}<0, \label{eq:q10}
\eeqn
and 
$$
\lim_{y\to \infty}Q(y)=\lim_{y\to \infty}\frac{c}{y^3}
\left\{\frac{y}{1+y^2}-\tan^{-1}y\right\}=0,
$$
so $-\frac{2c}{3}<Q(y)<0$ for all $y>0$.

Hence, from Eq. (\ref{eq:q10}) and Eq. (\ref{eq:n2}), we have 
\beqn
\lim_{y\to 0}h(y)&=&-\lim_{y\to 0}P(y)+\lim_{y\to 0}Q(y)=-\frac{2m}{R^3}+n^2-\frac{2c}{3}  \non \\
&=& -\frac{7m}{4R^3}+\frac{c}{R}
\biggm|\frac{1}{(1+R^2)R}-\frac{\tan^{-1} R}{R^2}\biggm|
-\frac{2c}{3}
= -\frac{7m}{4R^3}+\left[|Q(R)|-\frac{2c}{3}\right]<0. \non 
\eeqn
Similarly, we have  $\lim_{y\to\infty} h(y)=
-\lim_{y\to \infty}P(y)+\lim_{y\to \infty}Q(y)=n^2>0$.
Because when $y>0$, $h(y)$ is a monotonic function, 
there exists a unique point $\bar{y_1}>0$ such that 
$h(\bar{y_1})=0$.

For the case when $y<0$, from Eq.(\ref{eq:dp1})-(\ref{eq:dq1}), 
we have $P'(y)>0$,$Q'(y)<0$, so $h(y)$ is a monotonically decreasing 
function. 
We also have  $\lim_{y\to 0}h(y) <0$ and $\lim_{y\to -\infty}h(y)= n^2 > 0$.
Thus, similarly, we find that there is  
a unique point $\bar{y_2}<0$ such that 
$h(\bar{y_2})=0$. 
Therefore, there are two equilibrium 
points on the line $x=0$, i.e. $y$-axis. $\Box$

In the classical restricted three-body problem, there
are three collinear points, $L1, L2, L3$  
and two triangular points, i.e. $L4, L5$.
Theorem 1 shows the existence and uniqueness
of L4 and L5 for the system we consider here.

We now investigate the existence of equilibrium 
points on the $x$-axis. 
For convenience, we define 
\beq
S(x)\equiv \frac{m(x+R)}{|x+R|^3}+\frac{m(x-R)}
{|x-R|^3}-n^2 x, \label{eq:p2}
\eeq
and therefore,
\beq
S(x) =\left\{
\begin{array} {lll}
&  \frac{m}{(x+R)^2}+
\frac{m}{(x-R)^2}-n^2x,  &\quad {\rm for} \,\, x>R, \\
& \frac{m}{(x+R)^2}-\frac{m}{(x-R)^2}-n^2x, & \quad {\rm for} \,\, -R<x<R,\\
& -\frac{m}{(x+R)^2}-
\frac{m}{(x-R)^2}-n^2x, & \quad {\rm for} \,\, x<-R.
\end{array}\right. \label{eq:p2_3}
\eeq

$S(x)$ includes the gravitational
forces from the central binary and also the centrifugal force. 
From Eq.(\ref{eq:kxe}) and Eq.(\ref{eq:p2_3}), we have 
\beq
k(x)= -S(x)+c\left\{\frac{1}
{(1+x^2)x}-\frac{\tan^{-1}(x)}{x^2}\right\}= -S(x)+f_g(x), 
\label{eq:ak}
\eeq
where $f_g$ is defined in Eq.(\ref{eq:fg}).
The equilibrium points on the $x$-axis $(x_e,0)$ satisfy 
$A(x_e,0)=k(x_e)=0$.
In addition, we define the critical mass 
\beq
M_c\equiv \frac{51\pi m}{8[2R^3-3\tan^{-1} R+3R/(1+R^2)]}, 
\label{eq:alpha_ast}
\eeq
and it is always positive as shown in Fig. 1.
We also define
\beq
g_2(R,x)\equiv \frac{1}{(R-x)^4}-\frac{1}{(R+x)^4}\label{eq:g2}
\eeq
and
\beq
g_3(x)\equiv \frac{2}{x^4(1+x^2)^3}   
[9x^5+8x^3+3x-3(1+x^2)^3 \tan^{-1}x].\label{eq:f2}
\eeq

From Fig. 2 , we find that there is a
positive $x_c$ such that $g_3(x_c)=0$, and $g_3(x)>0$ for $x\in (0,x_c)$.
Because $g_3(-x)= -g_3(x)$, we also have 
$g_3(-x_c)=0$ and  $g_3(x)<0$ for $x\in (-x_c,0)$.
The numerical value of $ x_c \sim 1.43872$ approximately.
As in Fig. 2, we find that
when $x_c<x<R$, there is a lower bound $g_{3c}$ for $g_3(x)$
such that $g_{3c} < g_3(x)<0$.
The numerical value of $g_{3c}\sim -0.057$ approximately.
Finally, we define
\beq
g_4(R)\equiv
\frac{1}{(R-x_c)^4}-\frac{1}{(R+x_c)^4}
+\frac{17g_{3c}}{8[2R^3-3\tan^{-1} R+3R/(1+R^2)]}.
\label{eq:g4}
\eeq 
As shown in Fig. 3, $g_4(R)>0$ for any $R$.
The above definitions of $M_c, g_2(R, x), g_3(x), 
x_c, g_{3c}, g_4(R)$,
and their properties will be employed in below three theorems.

In the following Theorem 2, the results show  
the existence of three equilibrium points, 
i.e. Lagrange Point 1, 2, and 3 (L1-L3).   
In Theorem 3, we show that 
when the total mass of 
the galaxy $M_g>M_c$, 
two more equilibrium points, i.e. Jiang-Yeh Point 1 (JY1) 
and Jiang-Yeh Point 2 (JY2), exist.    
In Theorem 4, the results show that when the total mass of 
the galaxy $M_g<M_c$, 
neither JY1 nor JY2 exist.
Therefore, Jiang-Yeh Points exist if and only if 
the total galactic mass is larger than the
critical mass.

For convenience, in the following proofs of Theorem 2, 3, and 4, 
we denote $R^{+}$ to represent that $x$ tends to $R$ from the 
right hand side and $R^{-}$ to represent that $x$ tends to $R$ from the 
left hand side. 

\no {\bf  Theorem 2: The Existence of Lagrange Points L1, L2, and L3}\\ 
{\bf (i)} There is an  $x_1>R$ and  an $x_2<-R$ such that 
$k(x_1)=0$ and $k(x_2)=0$. 
That is, on the $x$-axis, there is an equilibrium point 
in the region $(R,\infty)$, and there is an 
equilibrium point in the region $(-\infty, -R)$ \\
{\bf (ii)} The origin (0,0) is an equilibrium point, 
i.e. $k(0)=0$. 

\noindent 
{\bf (Proof):}

\no {\bf (i)} 
When $x>R$, from Eqs. (\ref{eq:p2_3})-(\ref{eq:ak}),  
$$
k(x)=-S(x)+f_g(x) 
= -\frac{m}{(x+R)^2}-\frac{m}{(x-R)^2}+n^2x
+c\left\{\frac{1}{(1+x^2)x}-\frac{\tan^{-1}(x)}{x^2}\right\}. $$

Thus, we have 
$\lim_{x\to R^{+}} k(x)=\lim_{x\to R^{+}}-S(x)+f_g(x) =-\infty$ and
$\lim_{x\to \infty} k(x)=\lim_{x\to \infty}-S(x)+f_g(x)=\infty$. 
Therefore,  there is a point $x_1>R$ such that $k(x_1)=0$.

When $x<-R$,
using the similar method, we have that there is a point $x_2<-R$ such that 
$k(x_2)=0$.

\no {\bf (ii)}
When $-R<x<R$, from Eq. (\ref{eq:p2_3}), we have 
\beq
S(x) =\frac{m}{(x+R)^2}-\frac{m}{(x-R)^2}-n^2x.
\label{eq:p2_n0}
\eeq
Thus, $S(0)=0$. Moreover, from Eq.(\ref{eq:ak}), 
we employ L'Hopital Rule and obtain, 
$$
\lim_{x\to 0}f_g(x)
=\lim_{x\to 0}c\left\{\frac{x-(1+x^2)\tan^{-1}x}
{(1+x^2)x^2}\right\}
=\lim_{x\to 0}\frac{-c\tan^{-1}x}{1+2x^2}=0.$$ 
Therefore, $$k(0)=S(0)+\lim_{x\to 0}f_g(x)=0$$ 
and the origin (0,0) is an equilibrium point. 
$\Box$

\no {\bf  Theorem 3: The Existence of Jiang-Yeh Points JY1 and JY2}\\ 
If $M_g>M_c$, where $M_c$ is the critical mass defined in Eq. 
(\ref{eq:alpha_ast}), then there is an $x_4\in (0,R)$ such that 
$k(x_4)=0$ and an $x_5\in (-R,0)$ such that $k(x_5)=0$. \\

\no {\bf (Proof):}
  
When $-R<x<R$, from Eq.(\ref{eq:fg}) and 
Eq.(\ref{eq:p2_n0}), we have 
\beq
S'(x) =-\frac{2m}{(x+R)^3}+\frac{2m}{(x-R)^3}-n^2\label{eq:dp2}
\eeq
\beq
{\rm and} \quad
f'_g(x)=2c\left\{\frac{-2x^3-x+(1+x^2)^2\tan^{-1} x}
{x^3(1+x^2)^2}\right\}.\label{eq:dq2}
\eeq
Thus,
\beq
S'(0)=-\frac{4m}{R^3}-n^2\label{eq:dS0} 
\eeq
and
\beqn 
\lim_{x\to 0}f'_g(x)
& &=\lim_{x\to 0}
2c\left\{\frac{-2x^3-x+(1+x^2)^2\tan^{-1} x}
{x^3(1+x^2)^2}\right\} \non \\
& &= 2c\lim_{x\to 0}\frac{-5x^2+4x(1+x^2)\tan^{-1} x}{x^2(1+x^2)(3+7x^2)}\non\\
& &=2c \lim_{x\to 0}\frac{-6x+4(1+3x^2)\tan^{-1}x}{6x+40x^3+42x^5} \non\\
& &=2c  \lim_{x\to 0} \frac{-6+24x\tan^{-1} x+4\frac{1+3x^2}{1+x^2}}
{6+120x^2+210x^4}\non\\
& &=-\frac{2c}{3}
\label{eq:pq2_0}
\eeqn 

From Eq.(\ref{eq:n2}), Eq.(\ref{eq:pq2_0}) and 
Eq.(\ref{eq:alpha_ast}), we have 
\beqn
k'(0)
&=&-S'(0)+f'_g(0) \non  \\
&=&\frac{4m}{R^3}+n^2 -\frac{2c}{3} \non \\
&=&\frac{17m}{4R^3}+\frac{c}{R}
\biggm|\frac{1}{(1+R^2)R}-\frac{\tan^{-1} R}{R^2}\biggm|
-\frac{2c}{3}\non \\
&=& \frac{1}{R^3}\left\{\frac{17m}{4}+\frac{2M_g}{\pi}\left[-\frac{2R^3}{3}
+\tan^{-1}R-\frac{R}{(1+R^2)}\right] \right\} \non \\ 
&=& \frac{1}{R^3}\left\{\frac{17m}{4}+\frac{2M_g}{\pi}
\left(-\frac{17\pi m}{8M_c}\right) \right\} \non \\
&=&\frac{17m}{4R^3M_c} (M_c - M_g), 
\label{eq:McMg}
\eeqn
where $M_c$ is defined in Eq. (\ref{eq:alpha_ast}).
Due to the condition $M_g > M_c$, we obtain $k'(0) < 0 $. 

We first consider the region with $0<x<R$. 
From Eq.(\ref{eq:fg}) and 
Eq. (\ref{eq:p2_n0}), we have 
$\lim_{x\to R^{-}}S(x)=-\infty$ and  $-\infty < f_g(R)<0$. 
Thus,
$$\lim_{x\to R^{-}} k(x)=\lim_{x\to R^{-}}-S(x)+f_g(x)=\infty.$$
Because  $k(0)=0$, $k'(0)<0$, and 
$\lim_{x\to R^{-}} k(x)=\infty$, as $x$ increases from
zero, $k(x)$ becomes negative first and needs to be 
positive when $x$ approaches to $R$. 
Therefore, there is a point $0<x_4<R$ such that $k(x_4)=0$. 

Similarly, for the region with $-R<x<0$, 
we have $\lim_{x\to -R^{+}} k(x)=-\infty$, $k'(0)<0$, 
and  $k(0)=0$. As $x$ increases from $-R$,
$k(x)$ is negative first and needs to be positive when
$x$ is very close to zero due to that $k(0)=0$
and $k'(0)<0$. Therefore, 
there is a point $x_5\in (-R,0)$ 
such that $k(x_5)=0$.$\Box$

{\bf  Theorem 4: The Nonexistence of Jiang-Yeh Points JY1 and JY2} \\
When $M_g<M_c$, there is no any
$x\in (-R,0)$ or $x\in (0,R)$ 
such that $k(x)=0$. \\

\no {\bf (Proof):} 
Because $M_g<M_c$, from Eq. (\ref{eq:McMg}), we have 
$$k'(0)=\frac{17m}{4R^3M_c} (M_c - M_g)>0.$$

Since we will need to calculate $k''(x)$
later and $k''(x)=-S''(x) +f''_g(x)$,  we calculate
$S''(x)$ and $f''_g(x)$ here.
When $-R<x<R$, from Eq. (\ref{eq:dp2})-(\ref{eq:dq2}), 
we have
$$S''(x) =\frac{6m}{(x+R)^4}-\frac{6m}{(x-R)^4}= -6m g_2(R,x).$$
As we show in Appendix A, $S''(x)>0$ for $x\in (-R,0)$
and $S''(x)<0$ for $x\in (0,R)$.

If there is an $x$ such that $x_c< x< R $ 
(for the case $R>x_c$), 
we have $-S''(x) > 6m g_2(R, x_c)$ 
as shown in Appendix B.

Moreover,
$$
f''_g(x)=2c\left\{\frac{9x^5+8x^3+3x-3(1+x^2)^3\tan^{-1} x}
{x^4(1+x^2)^3}\right\}= c g_3(x).
$$
where $g_3(x)$ is defined in Eq.(\ref{eq:f2}).
Thus, the behavior of $f''_g(x)$ is completely determined
by the properties of $g_3(x)$. For example, 
$f''_g(x_c)= c g_3(x_c)=0$. Moreover, for  $x>0$,
we have $g_3(x)> g_{3c}$, where $g_{3c}$ is the lower bound 
of $g_3(x)$. Thus, $f''_g(x)= c g_3(x) > c g_{3c}$ for 
any $x\in (0,R)$.   

\no {\bf Case 1:} If $R<x_c$, 
then $g_3(x)>0$ for 
any $x\in (0,R)$ and $g_3(x)<0$ for $x\in (-R,0)$.  
We thus have $f''_g(x)>0$ 
for any $x\in (0,R)$ and 
$f''_g(x)<0$ for $x\in (-R,0)$.

 We first consider  the $0<x<R$ region.  
Since  $k'(0)>0$ and $k''(x)=-S''(x)+f''_g(x)>0$ for any $x\in (0,R)$, 
we have $k'(x)>0$.
Because $k(0)=0$ and $k'(x)>0$ for $0<x<R$, there is no  
$x\in (0,R)$ such that $k(x)=0$.

Similarly, we have $k(0)=0$ and $k'(x)>0$ for $-R<x<0$, 
so there is no any  $x\in (-R,0)$ such that $k(x)=0$.

\no {\bf Case 2:} If $R>x_c$, it will be more complicated as 
the $x$ we consider can be either larger or smaller than $x_c$,
such that $g_3(x)$ and $f''_g(x)$ can be negative or positive
in the region $x\in (0,R)$.  

(1) When $x < x_c$, we have $0<x < x_c<R$ here. Thus, 
$S''(x) < 0$, $g_3(x)>0$, 
and $f''_g(x)>0$, so 
$k''(x)=-S''(x)+f''_g(x)>0$. 

(2) When $x = x_c$, we have $g_3(x)=0$ and thus $f''_g(x)=0$.
 Due to that $S''(x) < 0$, $k''(x)=-S''(x)+f''_g(x)>0$.

(3) When  $x > x_c$, we have 
$-S''(x) > 6m g_2(R, x_c)$, and  $f''_g(x)=c g_3(x) > c g_{3c} $.
Moreover, due to that 
$g_{3c}<0$ and $M_g<M_c$, we have $M_g g_{3c}>M_c g_{3c}$.
Therefore, as defined before $c=2 M_g/\pi$ and from the definition 
of $M_c$ in Eq. (\ref{eq:alpha_ast}), we obtain 
\beqn
k''(x)&=&-S''(x)+f''_g(x) > 6m g_2(R, x_c)+ \frac{2}{\pi} M_g g_{3c}. \non \\
&>&  6m g_2(R, x_c)+ \frac{2}{\pi} M_c g_{3c} \non \\
&=&6m \left\{
\frac{1}{(R-x_c)^4}-\frac{1}{(R+x_c)^4}+\frac{17g_{3c}}{8[2R^3-3\tan^{-1} R+3R/(1+R^2)]} \right\} \non \\ 
&=&6m g_4(R). \non 
\eeqn
Although $g_2(R, x_c)>0$ and  $g_{3c}<0$, 
since $g_4(R)>0$ as shown in Fig. 3, we obtain $k''(x)> 0$.
From (1), (2), and (3), we know that $k''(x)> 0$ for any $x\in (0,R)$, 
no matter $x>x_c$ or not.

Therefore, due to that $k'(0)>0$
and $k''(x)>0$ for any $x\in (0,R)$, we have $k'(x)>0$.
Since $k(0)=0$ and $k'(x)>0$, there is no  
$x\in (0,R)$ such that $k(x)=0$.

For $-R<x<0$, by a similar procedure,
it can be shown that there is no  
$x\in (-R,0)$ such that $k(x)=0$ under the same condition. 
This concludes our proof of non-existence of 
JY1 and JY2.$\Box$
 
\section{The Bifurcations and Zero-Velocity Curves}

In order to demonstrate the 
analytic results proved above,
we determine the locations of equilibrium points 
numerically by solving $k(x)=0$, and $h(y)=0$, 
and plot in Fig. 4-5. 
The locations of the equilibrium points on the x-axis with $m=1$ 
for different values of $M_g$ 
are in Fig.4 (Panel (a) is for $R=0.25$ and Panel (b) 
is for $R=0.5$)
and Fig.5 (Panel (a) is for $R=1$ and Panel (b) is for $R=2$).

The values of the critical mass $M_c$ depend on 
$m$ and $R$. For Fig.4(a) with $m=1$ and $R=0.25$, we find that $M_c=9118.55$. 
Due to that $M_c$ is very huge, the value of $M_g$ cannot be larger than $M_c$
and there are always three equilibrium points, i.e. Lagrange Points, 
on the $x-axis$ for whole range of $M_g$ in that plot.
For Fig.4(b) with $m=1$ and $R=0.5$, we find that $M_c=339.12$. 
In this panel, the plot shows that there are three equilibrium points,
i.e. Lagrange Points L1, L2, L3, when $M_g < M_c$ and bifurcate into five 
equilibrium points, i.e. Lagrange Points, L1, L2, L3, 
and Jiang-Yeh Points, JY1, JY2, when   $M_g < M_c$.
As expected, the location of JY1 is in $(0,R)$ and the location of JY2
is in $(-R,0)$ for this case with $R=0.5$.

For Fig.5(a) with $m=1$ and $R=1$,
we find that $M_c=17.51$. Similarly, three equilibrium points,
L1, L2, L3, bifurcate into five equilibrium points, L1, L2, L3, JY1, JY2,
when $M_g$ passes through the value of $M_c$.
Finally, in Fig. 5(b) with $m=1$ and $R=2$, 
we find that the critical mass $M_c=1.44$.
Given that this $M_c$ is small,  there are three equilibrium points on the 
very left part, and five equilibrium points for the most 
part of this plot.  

In Fig. 6, the zero-velocity curves, which are obtained through 
Eq.(\ref{eq:Jacobi}), of the case with $m=1$ and $R=1$ 
are presented.  
Fig. 6(a) is for $M_g=10$, Fig. 6(b) is for $M_g=20$,
Fig. 6(c) is for $M_g=30$, and Fig. 6(d) is for $M_g=100$.
The $+$ signs are the locations for Lagrange Points, 
and the squares indicate the Jiang-Yeh Points. 
These points are 
numerically determined by solving $k(x)=0$ and $h(y)=0$.
Fig. 6 shows that they are completely consistent with the 
locations of equilibrium points implied by zero-velocity curves.
Due to that the critical mass $M_c=17.51$, apparently,
only Lagrange Points exist in Fig. 6(a),
but both Lagrange Points (L1-L5) and Jiang-Yeh Points (JY1-JY2) 
exist in Fig. 6(b)-(d).  
The locations and the Jacobi integral $C_J$ of 
all equilibrium points shown in Fig. 6 are summarized in Table 1.

{\centerline {\bf Table 1. The Locatins and $C_J$ of 
Equilibrium Points}}
    \begin{center}
       \begin{tabular}{|c|c|c|c|c|c|c|c|}\hline
\multicolumn{8}{|c|}{$M_g=10$} \\ \hline
&\multicolumn{2}{c|}{L1}  &L2& L3 & L4&\multicolumn{2}{|c|}{L5}  
\\ \hline
$(x_e,y_e)$ & \multicolumn{2}{c|}{(0,0)} &(1.70,0)&(-1.70,0) &(0,1.15) & \multicolumn{2}{|c|}{(0,-1.15)} \\ \hline
$C_J$ & \multicolumn{2}{c|}{16.73}&\multicolumn{2}{|c|}{17.35} &\multicolumn{3}{c|}{14.82}  \\ \hline \hline
\multicolumn{8}{|c|}{$M_g=20$} \\ \hline
&L1 & L2 & L3 & L4& L5 & JY1  &JY2 \\ \hline
$(x_e,y_e)$ &  (0,0) &(1.57,0)&(-1.57,0) &(0,1.09) &(0,-1.09) & (0.18,0)&
(-0.18,0)\\ \hline
$C_J$ & 29.46 &\multicolumn{2}{|c|}{30.14} &\multicolumn{2}{|c|}{26.67}
&\multicolumn{2}{|c|}{29.45} \\ \hline \hline
\multicolumn{8}{|c|}{$M_g=30$} \\ \hline
& L1 & L2& L3 & L4 & L5 & JY1  &JY2 \\ \hline
$(x_e,y_e)$ &  (0,0) &(1.50,0)&(-1.50,0) &(0,1.06) &(0,-1.06) & (0.36,0)&
(-0.36,0)\\ \hline
$C_J$ & 42.19 &\multicolumn{2}{|c|}{42.65} &\multicolumn{2}{|c|}{38.50}
&\multicolumn{2}{|c|}{42.0} \\ \hline  \hline
\multicolumn{8}{|c|}{$M_g=100$} \\ \hline
& L1 & L2 & L3 & L4 &  L5 & JY1  &JY2 \\ \hline
$(x_e,y_e)$ &  (0,0) &(1.34,0)&(-1.34,0) &(0,1.02) &(0,-1.02) & (0.62,0)&
(-0.62,0)\\ \hline
$C_J$ & 131.32 &\multicolumn{2}{|c|}{128.15} &\multicolumn{2}{|c|}{121.24}
&\multicolumn{2}{|c|}{127.55} \\ \hline 
\end{tabular}
    \end{center}
   
\normalsize

\section{The Stability of Equilibrium  Points}

The equations of the considered system can be written as
\beq \left\{
\begin{array}{ll}
&  \frac{dx}{dt}=u,  \\
& \frac{dy}{dt}=v, \\
& \frac{du}{dt}=2nv +A(x,y),\\
& \frac{d v}{dt}=-2nu+B(x,y),
\end{array}\right. 
\eeq
where $A(x,y)$, $B(x,y)$ are defined in Eqs.(\ref{eq:gf2})-(\ref{eq:gg2}).

To study the stability of equilibrium points, we need to know the
properties of eigen-values of equilibrium points.
The characteristic 
equation of the eigen-value $\lambda$ is  
\beq
\lambda^4+(4n^2-A_x-B_y)\lambda^2+2n(A_y-B_x)
\lambda+A_xB_y-B_xA_y=0, 
\label{eq:labd1}
\eeq
where $A_x\equiv {\partial A(x,y)}/{\partial x}$,
$A_y\equiv {\partial A(x,y)}/{\partial y}$,
$B_x\equiv {\partial B(x,y)}/{\partial x}$, and
$B_y\equiv {\partial B(x,y)}/{\partial y}$.
Thus,
\beqn
&&A_x=n^2-\frac{m}{r_1^3}-
\frac{m}{r_2^3}+\frac{3m(x+R)^2}{r_1^5}+\frac{3m(x-R)^2}{r_2^5}+\frac{c}{r}
 \left\{\frac{1}{(1+r^2)r}-\frac{\tan^{-1} (r)}{r^2}\right\} \non \\
& & \qquad +\frac{cx}{r}\left\{\frac{-5r^3-3r+3(1+r^2)^2\tan^{-1} r)}{(1+r^2)^2r^3}\right\},\label{eq:A_x} \\
& & A_y=\frac{3m(x+R)y}{r_1^5}+\frac{3m(x-R)y}{r_2^5}+
\frac{cxy(-5r^3-3r+3(1+r^2)^2\tan^{-1} r)}{(1+r^2)^2r^5}, \label{eq:A_y} \\
& & B_x=\frac{3my(x+R)}{r_1^5}+\frac{3my(x-R)}{r_2^5}+\frac{cxy(-5r^3-3r+3(1+r^2)^2\tan^{-1} r)}{(1+r^2)^2r^5},\label{eq:B_x}\\
& & B_y=n^2-\frac{m}{r_1^3}-\frac{m}{r_2^3}+\frac{3my^2}{r_1^5}
+\frac{3my^2}{r_2^5}+\frac{c}{r}\left\{\frac{1}{(1+r^2)r}-\frac{\tan^{-1} (r)}{r^2}\right\} \non \\
&& \qquad +\frac{cy^2(-5r^3-3r+3(1+r^2)^2\tan^{-1} r)}{(1+r^2)^2r^5}.\label{eq:B_y}
\eeqn
From Eq.(\ref{eq:A_y}) and Eq.(\ref{eq:B_x}), 
for any $x_e$ we have 
$A_y(x_e,0)=B_x(x_e,0)=0$, 
and for any $y_e$ we have 
$A_y(0,y_e)=B_x(0,y_e)=0$.
 
In order to do further investigation,
some parameters need to be specified and 
we set $m=1$ and $R=1$ for all the results in this section.
Thus, from Eq.(\ref{eq:fg}), Eq.(\ref{eq:n2}), and Eq.(\ref{eq:alpha_ast})  
we have
$$n^2=\frac{1}{4}+\frac{2M_g}{\pi}\left(\frac{\pi}{4}-\frac{1}{2}\right),
$$
and 
$$
M_c=\frac{51\pi}{28-6\pi}\sim 17.5.
$$

At first, we consider the equilibrium point L2 or JY1,
$(x_e,y_e)$, which satisfies $k(x_e)=0$ with $x_e>0$ and $y_e=0$.
Because $A_y(x_e,0)=0$ and $B_x(x_e,0)=0$, 
Eq.(\ref{eq:labd1}) becomes 
\beq
\lambda^4+(4n^2-A_x-B_y)\lambda^2+A_x B_y=0. 
\label{eq:labd2}
\eeq

For convenience, we define $\Omega=A_xB_y$ and
$\Pi\equiv A_x+B_y-4n^2$
Therefore, we have
\beq
\lambda^2_{+}=\frac{\Pi+\sqrt{\Pi^2-4\Omega}}{2} \quad {\rm and} \quad 
\lambda^2_{-}=\frac{\Pi-\sqrt{\Pi^2-4\Omega}}{2}.
\label{eq:lamda3}
\eeq

Moreover, $A_x(x_e,0)$ and  $B_y(x_e,0)$ can be 
expressed as:
\beqn
& &A_x(x_e,0)=3n^2-\frac{2}{x_e}\left(\frac{1}{|x_e+1|^3}
-\frac{1}{|x_e-1|^3}\right)-\frac{4M_g}{\pi(1+x_e^2)^2}  \non \\
&&=\frac{3}{4}+\frac{6M_g}{\pi}\left(\frac{\pi}{4}-\frac{1}{2}\right)-
\frac{2}{x_e}\left(\frac{1}{|x_e+1|^3}
-\frac{1}{|x_e-1|^3}\right)-\frac{4M_g}{\pi(1+x_e^2)^2},
\label{eq:Ax_xe} \\
&& B_y(x_e,0)=\frac{1}{x_e}\left(\frac{1}{|{x_e}+1|^3}
-\frac{1}{|{x_e}-1|^3} 
\right) <0.
\label{eq:By_xe}
\eeqn

For the equilibrium point L2, $x_e>R=1$, so $(1+x_e^2)^2>4$. 
Thus, 
\beq
 -\frac{4M_g}{\pi(1+x_e^2)^2}>-\frac{M_g}{\pi}. \label{eq:test1}
\eeq
Because $x_e>1$ leads to 
$$
\frac{3}{4}
-\frac{2}{x_e}\left(\frac{1}{|x_e+1|^3}
-\frac{1}{|x_e-1|^3}\right) > 0,
$$
with Eq.(\ref{eq:Ax_xe}) and Eq.(\ref{eq:test1}), we obtain 
$$ A_x(x_e,0)> \frac{6M_g}{\pi}\left(\frac{\pi}{4}-\frac{1}{2}\right)
-\frac{4M_g}{\pi(1+x_e^2)^2}>\frac{M_g}{\pi}\left(\frac{3\pi}{2}-4\right)>0. $$ 
Since $A_x(x_e,0)>0$ and $B_y(x_e,0)<0$, we have 
$\Omega=A_x(x_e,0)B_y(x_e,0)<0$. 
Thus, $\Pi^2-4\Omega>0$, and we have $\lambda^2_{+}>0$
and $\lambda^2_{-}<0$. 
As in Szebehely (1967), 
this result indicates that it is an unstable equilibrium point.  

For the equilibrium point JY1 (which we assume it exists with
$M_g>M_c$), it has $0<x_e<1$ and $y_e=0$.   
For a given $M_g$, we numerically search the location of 
JY1, and then obtain the corresponding 
 $A_x(x_e,0)$.
The numerical value of $A_x(x_e,0)$ as a function of 
$M_g$ is shown in Fig. 7(a).  It shows that $A_x(x_e, 0)>0$
for the considered $M_g$.
From Eq.(\ref{eq:By_xe}), we have $B_y(x_e,0)<0$ here
and so $\Omega=A_xB_y<0$. 
Thus, $\Pi^2-4\Omega>0$, we have $\lambda^2_{+}>0$
and $\lambda^2_{-}<0$. 
Therefore, JY1 is also an unstable equilibrium point.
Because our system is symmetric with respect to the 
$y$-axis, the above results imply that the equilibrium 
point L3 and JY2 are unstable.

Secondly, we study the equilibrium point L4, which can be written as 
$(0, y_e)$ with $y_e > 0$. 
As mentioned previously,
we have $A_y(0,y_e)=B_x(0,y_e)=0$. Thus, 
Eq.(\ref{eq:labd2}) and Eq.(\ref{eq:lamda3})
are also valid here.
We find that
\beq
A_x(0,y_e)=\frac{6}{(1+y_e^2)^{3/2}}>0,
\eeq
but cannot determine the sign of $B_y(0, y_e)$
analytically. Thus, for each given $M_g$, 
the location of L4, $(0, y_e)$, and then the 
corresponding value of $\Pi^2-4\Omega$ is 
determined numerically. For $M_g < 100$,
we find that $\Pi^2-4\Omega < 0$ as shown in Fig. 7(b).
Thus, both $\lambda^2_{+}$
and $\lambda^2_{-}$ are complex numbers. 
This leads to that both $\lambda_{+}$ and $\lambda_{-}$ 
have a root which contains positive real part, so that 
L4 is unstable. 
As our system is symmetric with respect to the $x$-axis,
the equilibrium point L5 is therefore unstable. 

Thirdly, we study the equilibrium point L1, (0,0), here.
As $A_x, A_y, B_x, B_y$ contain terms that both denominator and numerator 
become zero
at $(x,y)=(0,0)$, we consider the limit that $r$ approaches to zero
and employ L'Hopital Rule for the following results 
(see Appendix C for the details).  
We obtain $A_y(0,0)=0$ and $B_x(0,0)=0$,     
so Eq.(\ref{eq:labd2}) and Eq.(\ref{eq:lamda3}) are also valid for L1 here.  
In addition, we have $B_y(0,0)<0$, and
\beq
A_x(0,0)=\frac{17}{4M_c} (M_c-M_g).
\eeq

When $M_g < M_c$, $A_x(0,0)>0$ and thus $\Omega < 0$.
Therefore, $\Pi^2 -4\Omega >0$, and we have
$\lambda^2_{+}>0$, $\lambda^2_{-}<0$. This leads to an unstable point.
When $M_g > M_c$, $A_x(0,0)<0$ and thus $\Omega > 0$.
The sign of  $\Pi^2 -4\Omega$ cannot be determined analytically.
It is shown in Fig. 7(c) numerically that $\Pi^2 -4\Omega>0$.
Further, we can show that (see Appendix C)
\beq
\Pi=  \frac{3}{2} -\frac{2}{\pi}M_g(\frac{\pi}{2}+\frac{1}{3}).
\eeq 
Since $\Pi$ has a simple linear dependence on $M_g$, it is trivial that 
$\Pi < 0$ for the considered value of $M_g$ here.
This leads to that both $\lambda^2_{+}$ and $\lambda^2_{-}$
are real and negative.
Thus, both $\lambda_{+}$ and $\lambda_{-}$ are pure imaginary numbers.
Therefore, L1 is a center for this case.

In this section, we have shown that (1)
when $M_g < M_c$, there are five equilibrium points in total, 
i.e. L1, L2, L3, L4, L5, and all of them are unstable;
(2) when $M_g > M_c$, there are seven equilibrium points in total,
and among these, L2, L3, L4, L5, JY1, JY2 are unstable, but L1 is neutrally
stable. 

\section{The Summary}

We have studied the equilibrium points of a galactic system  
with a supermassive binary black hole.
Focusing on 
the case with an equal mass binary black hole,  
the conditions of existence of equilibrium points,
including Lagrange Points and Jiang-Yeh Points,
are investigated. A necessary and sufficient
condition for the existence of Jiang-Yeh Points is found. 
That is, Jiang-Yeh Points exist  
if and only if the total galactic mass is larger than the
critical mass.
The further stability analysis shows that  Jiang-Yeh Points
are unstable. 
Thus, many stars might be ejected from the region near 
Jiang-Yeh Points after a supermassive binary black hole
come to near the core of a galaxy during merging events.
Therefore, the unstable Jiang-Yeh Points could lead to 
a mechanism to form galactic cores, 
as those observed in Lauer et al. (2007).
Although N-body simulations are needed to confirm this possible
explanation about the core formation of galaxies,
we conclude that our model and the results are 
fundamentally important for the study of 
galactic structures.
 
\section*{Acknowledgment}
We thank the referee for
very good suggestions which greatly improved this paper.
We are grateful to the National Center for High-performance Computing
for computer time and facilities. 
This work is supported in part 
by the National Science Council, Taiwan, under 
Ing-Guey Jiang's
Grants NSC 100-2112-M-007-003-MY3
and Li-Chin Yeh's 
Grants NSC 100-2115-M-134-004.

\clearpage
\cl{\bf Appendix A}
\no
Lemma 1:     $S''(x) >0$ for  $-R<x<0$ and
     $S''(x) <0$ for $0<x<R$ 

\no {\bf (Proof):} \\
If $-R<x<0$, then  $|x-R|>|x+R|$ and
$(x-R)^4>(x+R)^4$. Thus,  $S''(x)=6m\left\{\frac{1}{(x+R)^4}-\frac{1}
{(x-R)^4}\right\}>0.$ On the other hand, if $0<x<R$, then $|x+R|>|x-R|$ and
$(x+R)^4>(x-R)^4$. Therefore,
 $S''(x)=6m\left\{\frac{1}{(x+R)^4}-\frac{1}{(x-R)^4}\right\}<0.$ $\Box$
\\
\\
\cl {\bf Appendix B}
\no
Lemma 2: If $x_c<x<R$, then $ -S''(x)> 6m g_2(R,x_c).$

\no {\bf (Proof):}\\
If $x_c<x<R$, then $|x_c-R|>|x-R|$ and $|x+R|>|x_c+R|$. Moreover, we have
$(x_c-R)^4>(x-R)^4$ and $(x+R)^4>(x_c+R)^4$. Thus, From Eq. (\ref{eq:g2}),
$$S''(x)=6m\left\{\frac{1}{(x+R)^4}-\frac{1}
{(x-R)^4}\right\}<6m\left\{\frac{1}{(x_c+R)^4}-\frac{1}
{(x_c-R)^4}\right\}= -6m g_2(R,x_c),$$
so $-S''(x)>6m g_2(R,x_c).$ $\Box$
\\
\\
\cl {\bf Appendix C}
The calculations of  
$A_x(0,0), A_y(0,0), B_x(0,0), B_y(0,0)$ are presented here.
The results here were used in the section of stability analysis. 

For $A_x(0,0)$, we first consider the summation of 2nd to 
5th terms. Because we set $m=1$ and $R=1$ for the results in the section 
of stability analysis, $r_1=r_2=1$ and thus
\beq
\lim_{r\to 0} -\frac{m}{r_1^3}-\frac{m}{r_2^3}
+\frac{3m(x+R)^2}{r_1^5}+\frac{3m(x-R)^2}{r_2^5}=4.
\eeq
We then apply the L'Hopital Rule on the 6th term of 
Eq. (\ref{eq:A_x}) as 
\beqn
&&\lim_{r\to 0}\frac{c}{r} \left\{\frac{1}{(1+r^2)r}
-\frac{\tan^{-1} (r)}{r^2}\right\} \non \\
&&=\lim_{r\to 0}\frac{c[r-(1+r^2)\tan^{-1}r]}{(1+r^2)r^3} \non \\
&&=\lim_{r\to 0}\frac{c[1-1-2r\tan^{-1}r]}{5r^4+3r^2} \non \\
&&=\lim_{r\to 0}\frac{-2c\tan^{-1}r}{5r^3+3r} \non \\ 
&&=\lim_{r\to 0}\frac{-2c\frac{1}{1+r^2}}{15r^2+3} \non \\
&&=-\frac{2c}{3} \label{eq:fover_r}
\eeqn
Similarly, the 7th term becomes
\beqn
&&\lim_{r\to 0} \frac{cx}{r} \left\{\frac{-5r^3-3r+3(1+r^2)^2
\tan^{-1} r}{(1+r^2)^2r^3}\right\} \non \\
&&=\lim_{r\to 0} \frac{cx}{r}
\lim_{r\to 0} \left\{\frac{-5r^3-3r+3(1+r^2)^2
\tan^{-1} r}{(1+r^2)^2r^3}\right\} \non \\
&&=c\lim_{r\to 0}\frac{-12r^2+12r(1+r^2)\tan^{-1}r}
{3r^2+10r^4+7r^6} \non \\
&&=c\lim_{r\to 0}\frac{-24r+12r+12(1+3r^2)\tan^{-1}r}
{6r+40r^3+42r^5} \non \\
&&=c\lim_{r\to 0}
\frac{-12+12\frac{1+3r^2}{1+r^2}+72r\tan^{-1}r}{6+120r^2+210r^4}\non \\
&&=0.\label{eq:df_r}
\eeqn
From the above results and $c=2M_g/\pi$,
\beqn
A_x(0,0)&=&n^2+4-\frac{2c}{3} \non \\
&=& n^2+4-\frac{4}{3\pi}M_g \non \\
&=&\frac{17}{4}+\frac{2M_g}{\pi}\left(\frac{\pi}{4}-\frac{1}{2}\right)
-\frac{4M_g}{3\pi} \non \\
&=&\frac{17}{4}+\left(\frac{1}{2}-\frac{7}{3\pi}\right)M_g \non \\
&=&\frac{17}{4M_c}(M_c-M_g).\label{eq:Ax_0}
\eeqn

Through the similar process, we can easily show that 
$A_y(0,0)=0$, $B_x(0,0)=0$ and also obtain that
\beqn
B_y(0,0)&=&n^2-2-\frac{2c}{3} \non \\
&=&n^2-2-\frac{4M_g}{3\pi} \non \\
&=&-\frac{7}{4}-\frac{4}{3\pi}M_g
+\frac{2M_g}{\pi}\left(\frac{\pi}{4}-\frac{1}{2}\right)  \non \\
&=&-\frac{7}{4}+\frac{2M_g}{\pi}\left(\frac{\pi}{4}
-\frac{7}{6}\right)  \non \\
&<&0.\label{eq:By_0}
\eeqn
The corresponding $\Pi$  is defined as
\beqn
\Pi&\equiv& A_x(0,0)+B_y(0,0)-4n^2\non \\
&=&-2n^2+2-\frac{8}{3\pi}M_g\non \\
&=&\frac{3}{2}-\frac{2}{\pi}M_g\left(\frac{\pi}{2}+\frac{1}{3}\right),
\eeqn
which is a linear function of $M_g$.

\clearpage
\begin{figure}[ht]
\centering
\includegraphics[width=0.5\textwidth]{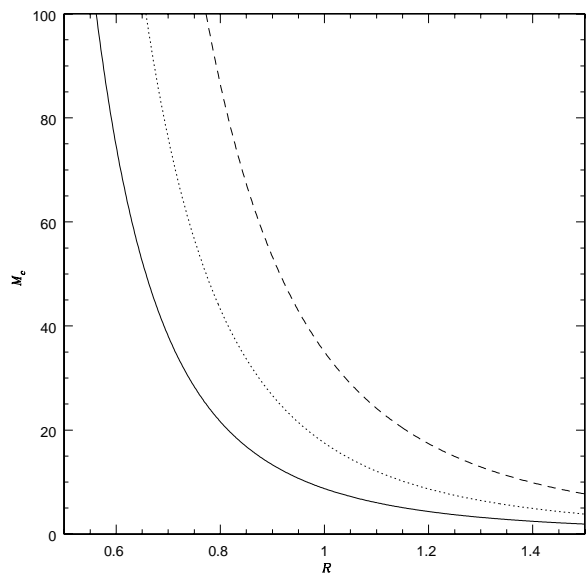}
\caption{The behavior of $M_c$ as a function of $R$.
The solid curve is for $m=0.5$, the dotted curve is for
$m=1$, and the dashed curve is for $m=2$.} \label{fig:Mc}
\end{figure} 

\clearpage
\begin{figure}[ht]
\centering
\includegraphics[width=0.5\textwidth]{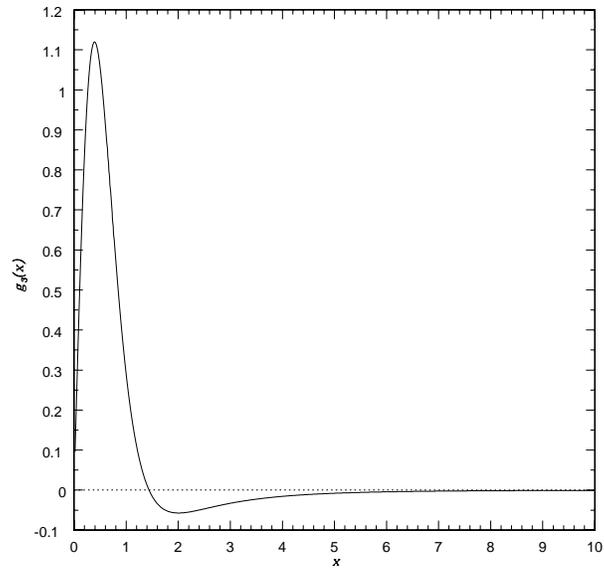}
\caption{The behavior of the function $g_3(x)$.} \label{fig:g3}
\end{figure}

\clearpage
\begin{figure}[ht]
\centering
\includegraphics[width=0.5\textwidth]{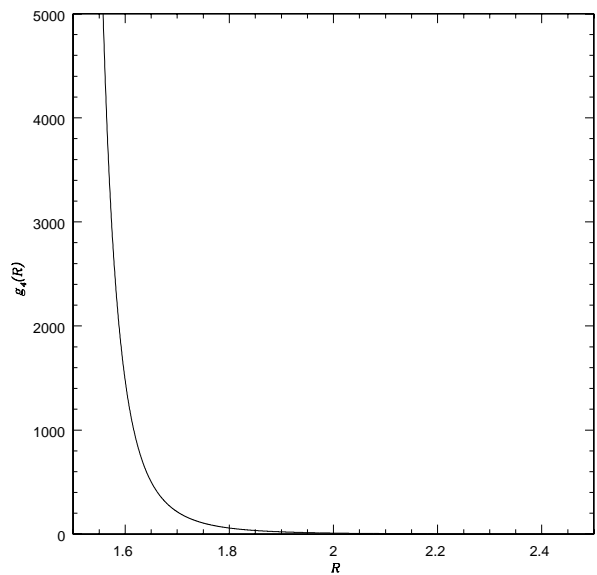}
\caption{The behavior of the function $g_4(R)$.} \label{fig:g4}
\end{figure}

\clearpage
\begin{figure}[ht]
\centering
\includegraphics[width=0.75\textwidth]{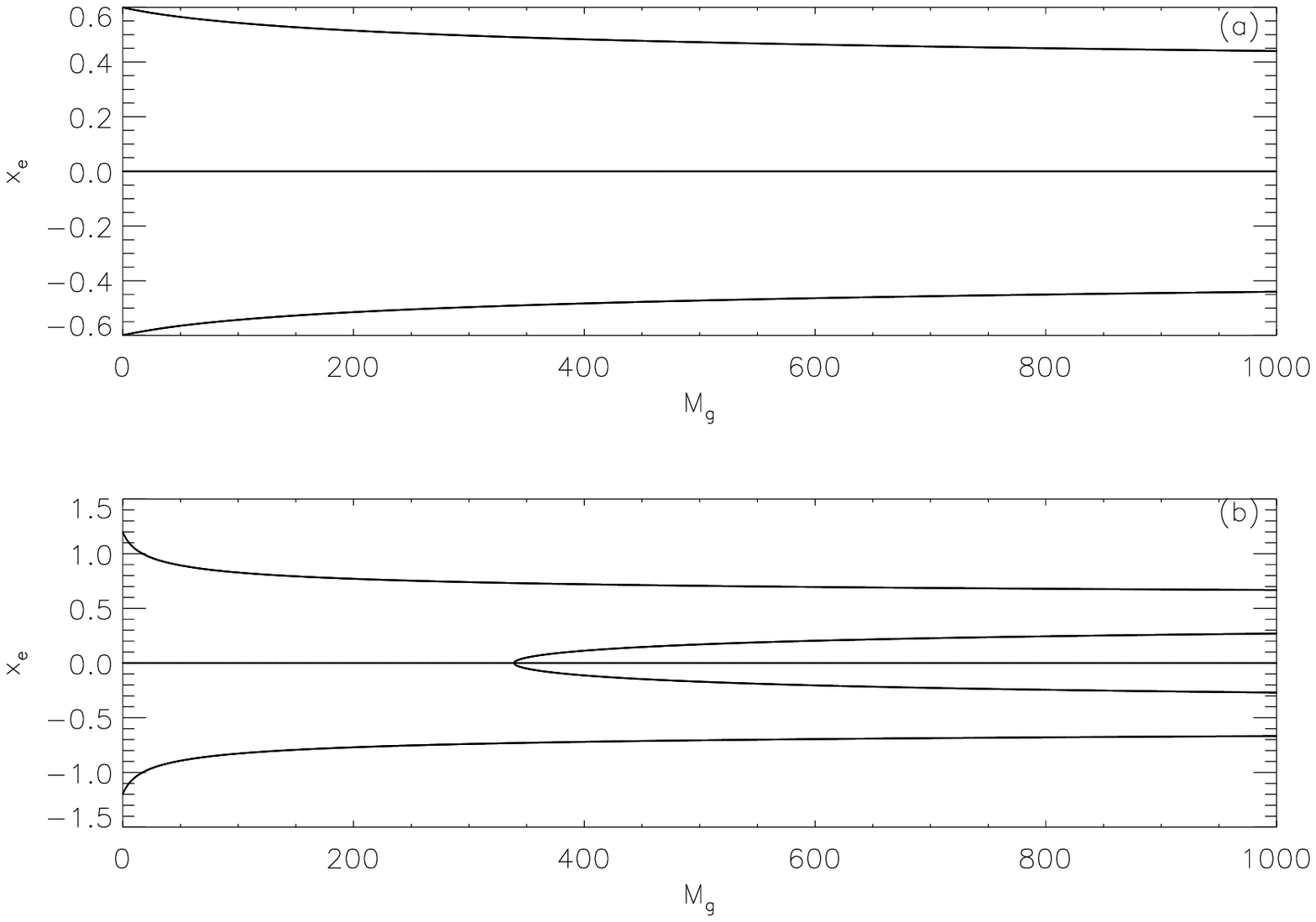}
\caption{The locations of the equilibrium points on the $x$-axis with $m=1$ 
as a function of $M_g$. (a) is for $R=0.25$ and (b) is for $R=0.5$.} 
\label{fig:fig1}
\end{figure}

\clearpage
\begin{figure}[ht]
\centering
\includegraphics[width=0.75\textwidth]{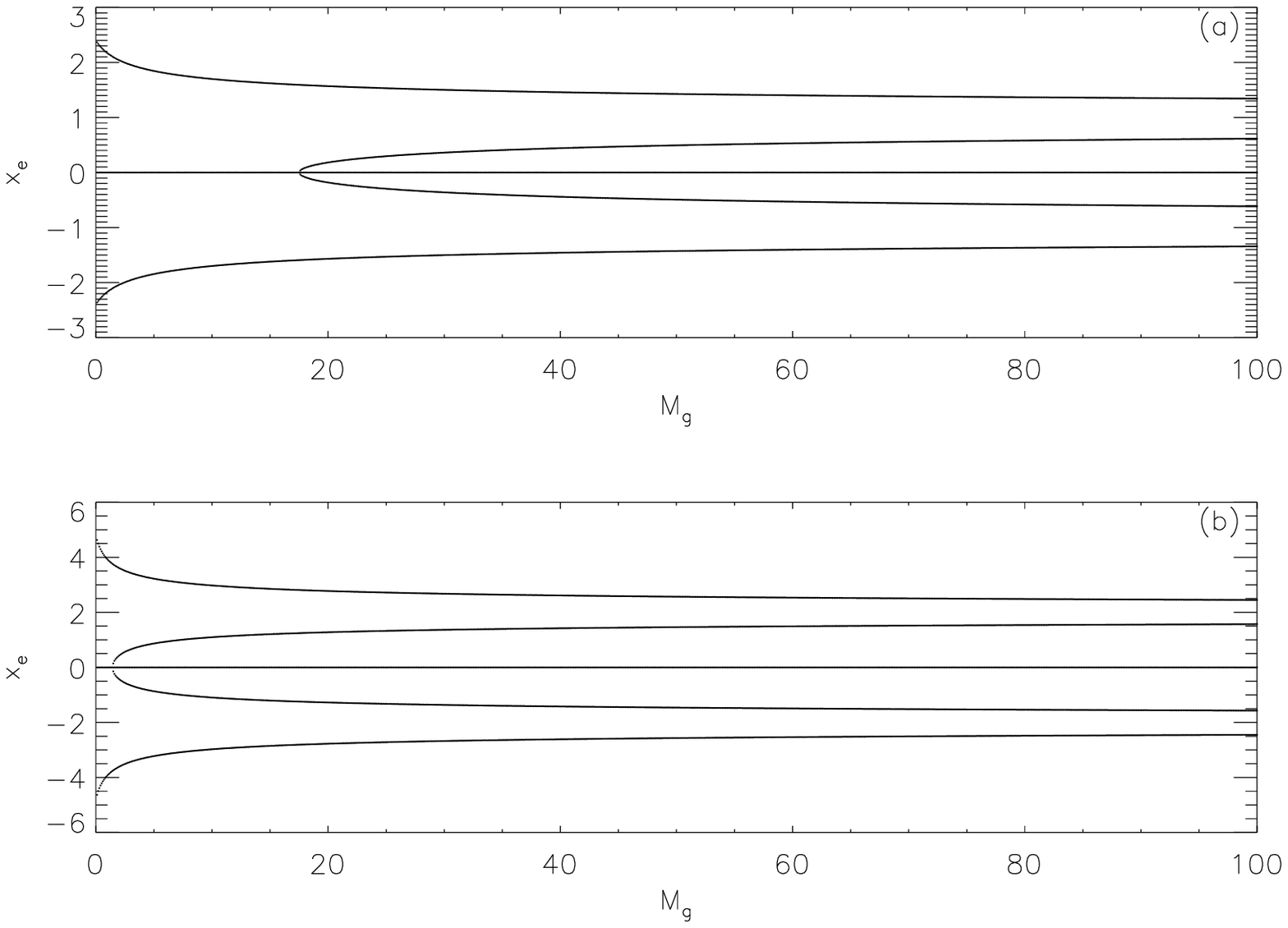}
\caption{The locations of the equilibrium points on the $x$-axis with $m=1$ 
as a function of $M_g$. (a) is for $R=1$ and (b) is for $R=2$.} 
 \label{fig:figx_r12}
\end{figure}

\clearpage
\begin{figure}[ht]
\centering
\includegraphics[width=1.\textwidth]{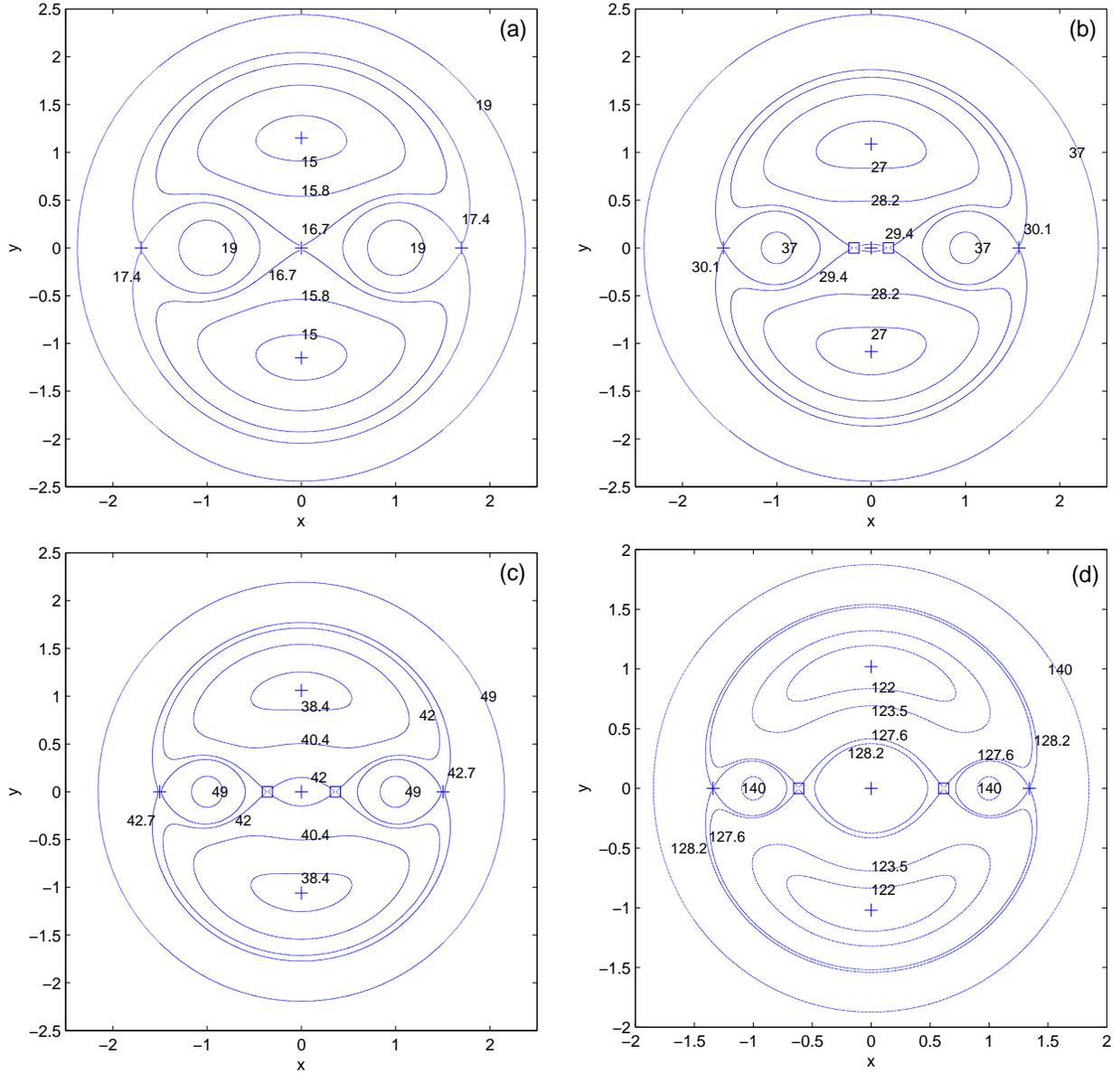}
\caption{The zero-velocity curves of the system when $R=1$ and $m=1$,
on which  
the corresponding values of the Jacobi integral $C_J$ are labelled.  
(a) is for $M_g=10$, (b) is for $M_g=20$,
(c) is for $M_g=30$, and (d) is for $M_g=100$.
The $+$ signs indicate the locations of Lagrange Points
and the squares indicate the locations of Jiang-Yeh Points.
}\label{fig:fig6_all}
\end{figure}

\clearpage
\begin{figure}[ht]
\centering
\includegraphics[width=1.\textwidth]{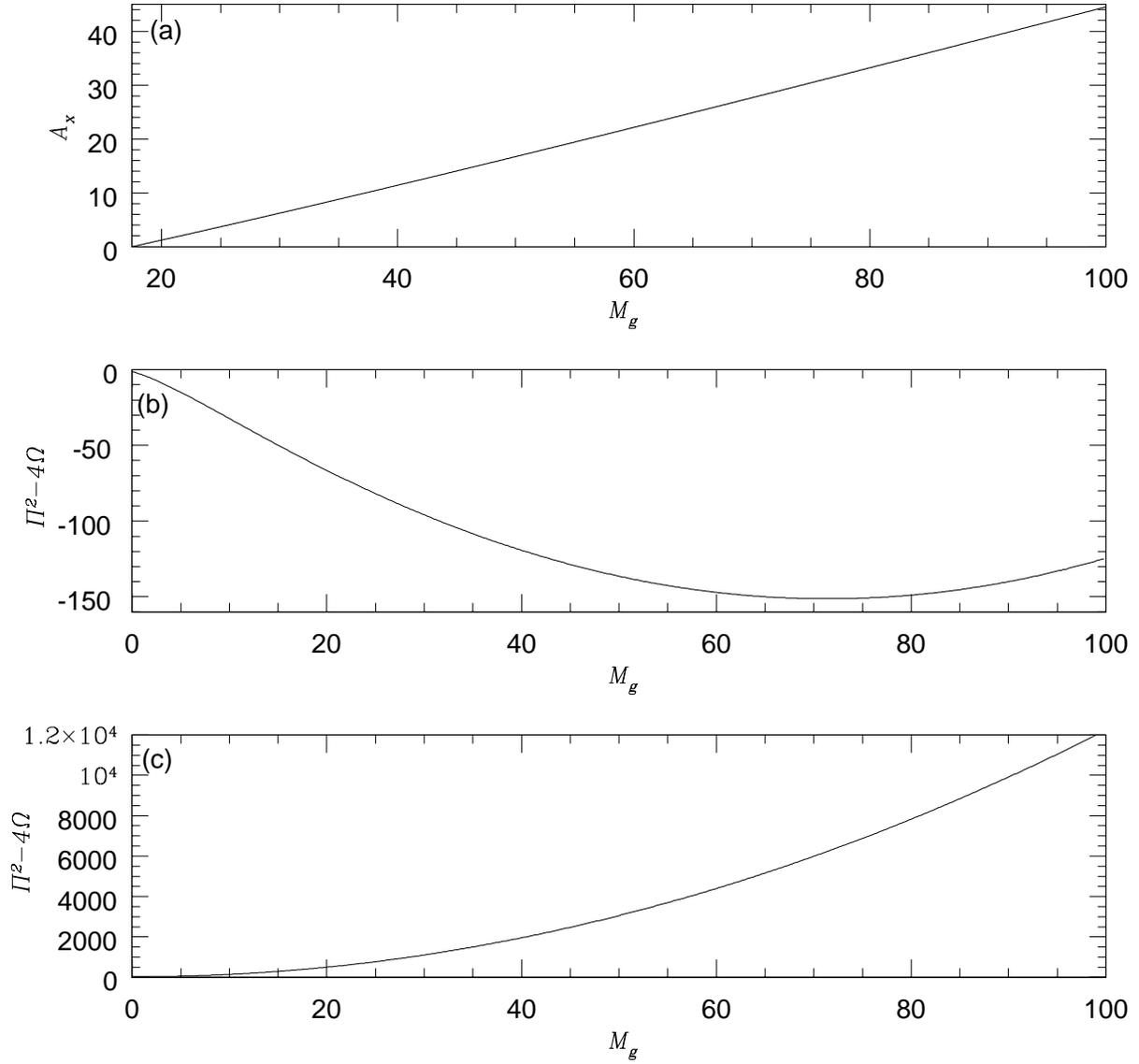}
\caption{(a) The value of $A_x$ as a function of $M_g$ for JY1.
(b) $\Pi^2 - 4 \Omega$ as a function of $M_g$ for L4. 
(c) $\Pi^2 - 4 \Omega$ as a function of $M_g$ for L1
    when $M_g > M_c$.
}\label{fig:fig7}
\end{figure}

\end{document}